\documentstyle[12pt]{article}
\begin{document}
\baselineskip 9 mm
\date{ }
\title{ \bf Bubbling  and  bistability in two  parameter  discrete  systems}
\vspace {1 cm}
\author {G Ambika and Sujatha N V\\
Department of Physics\\
Maharajas College\\
Cochin-682 011\\
INDIA.}
\maketitle
\vspace{2 cm}

\begin{abstract}
We present a graphical analysis of the mechanisms underlying the 
occurrences of bubbling sequences and bistability regions in the bifurcation 
scenario of a special class of one dimensional two parameter maps. The main 
result of the analysis is that whether it is bubbling or bistability is 
decided by the sign of the third derivative at the inflection point of the 
map function.
\end{abstract}
\hspace{1.1 cm}
PACS numbers: 05.45. +b, 05.40.+j

\pagebreak
{\large \bf {1. Introduction.}}\\
The studies related to onset of chaos in one- dimensional discrete systems 
modeled by nonlinear maps, have been quite intense and exhaustive during the 
last two decades. Such a system normally supports a sequence of period 
doublings leading to chaos. It is also possible to take it back to 
periodicity through a sequence of period halvings by adding perturbations 
or modulations to the original system[1,2]. This has, most often, been 
reported as a mechanism for control of chaos. In addition, there are 
features like tangent bifurcations, intermittency, crises etc., that occur 
inside the chaotic regime and are not of immediate relevance to the present 
work. However, if the system is sufficiently nonlinear, there are other 
interesting phenomena like bubble structures and bistability that have 
invited comparatively less attention. The simplest cases where these are 
realised are maps with at least two control parameters, one that controls 
the nonlinearity and the other which is a constant additive one. i.e., maps of 
the type,  
\begin{equation}
 X_{n+1} = f (X_n, a, b) = f_1 (X_n, a)+ b                          
 \end{equation}
 In these maps, if $a$ is varied for a given $b$, the usual period doubling 
 route to chaos is observed. But when $a$ is kept at a point beyond the first 
 period doubling point $a_1$, and $b$ is varied, the first period doubling is 
 followed by a period halving forming a closed loop like structure called 
 the primary bubble in the bifurcation diagram(Fig(1.a)). If $a$ is kept beyond the 
 second bifurcation point $a_2$, and $b$ is tuned, secondary bubbles appear on 
 the arms of the primary bubble. Thus as we shift the map along the $a$-axis 
 and drift it along the $b$-axis, the complete bubbling scenario develops in 
 the different slices of the space $(X,a,b)$. This accumulates into what is 
 known as bimodal chaos- chaos restricted or confined to the arms of the 
 primary bubble. This can be viewed as a separate scenario to chaos in such 
 systems.\\
 It has been confirmed that the Feigenbaum indices for this scenario with 
 $a$ as control parameter would be the same as the $\alpha$ and $\delta$ of 
 the normal period doubling route to chaos[3]. However, detailed RG analysis
 by Oppo and Politi[4], involving the parameter $b$ also, indicates that if
 $b$ is kept at a critical value, $b_{c}$, where bimodal chaos just disappears, 
 then there is a slowing down in the convergence rate leading to an index 
 which is $(\delta)^{1/2}$. This has been experimentally verified in a
 $CO_{2}$ laser system with modulated losses[5]. 
 The bubbling scenario is seen in the bifurcation diagrams of many nonlinear 
 systems like coupled driven oscillators[6, 7], oscillatory chemical 
reactions, diode circuits, lasers[8,9], insect populations[10], cardiac 
cell simulations[11], coupled or modulated maps[12,13], quasi-periodically 
forced systems[14], DPCM transmission system[15] and traffic flow 
systems[16] etc. The very fact that this phenomenon appears in such a wide 
variety of systems makes it highly relevant to investigate and expose the 
common factor(s) in them i.e., the underlying basic features that make them 
support bubbles in their bifurcation scenario. The above mentioned 
continuous systems require maps with at least two parameters of type (1) to 
model them, the second additive parameter being the coupling strength, 
secondary forcing amplitude etc. We note that in all the above referred 
papers no specific mention is made regarding the mechanism of formation of 
bubbles, probably because the authors were addressing other aspects of the 
problem. However, there have been a number of isolated attempts to analyse 
the criteria for bubble formation in a few typical systems. According to 
Bier and Bountis[3], the two criteria are, the map must possess some 
symmetry and the first period doubling should occur facing the symmetry 
line. Later, Stone[17] makes these a little more explicit by stating that 
the map should have an extending tail (with a consequent inflection point) 
and the inflection point should occur to the right of the critical point of 
the map. It is clear that this applies only to maps with one critical point. 
The relation of the extending tail to bubbling is briefly discussed in [18] 
also.\\
Bistability is an equally interesting and common feature associated with 
many nonlinear systems like a ring laser[19] and a variety of electronic 
circuits[20] .A recent renewal of interest in such systems arises from the 
fact that they form ideal candidates for studies related to stochastic 
resonance phenomena[21]. The bistable behaviour in two parameter maps 
is shown in the bifurcation diagram in Fig(1.b). To the best of 
our knowledge, attempts to study any type of conditions for the occurrence 
of  bistability are so far not seen reported in the literature.\\ 
Our motivation in the present work is to generalise the criteria reported 
earlier for bubbling and put them together with more clarity and simplicity. 
As a byproduct, we succeed in stating the conditions for bistability also 
along similar lines in systems of type (1), even though these are two 
mutually exclusive phenomena as far as their occurrence regime is concerned. 
We provide a detailed graphical analysis, which leads to a simple and  
comprehensible  explanation  for  the  same.\\ 
The paper is organized as follows. In section 2, the criteria for 
bistability and bubbling are stated followed by a brief explanation. 
The graphical analysis taking two simple cubic maps as examples is included 
in section 3 and the concluding comments in section 4.\\
\pagebreak

{\large \bf {2. The dynamics of bubbling and bistability}}\\
For the special class of maps given in (1), the basic criteria for 
bubbling / bistability can be stated as follows.\\
   {\it {The non-linearity in $f(X, a, b)$ must be more than quadratic.}}\\
This implies that, $f\ ^{\prime}(X, a, b)$ (the prime indicating derivative 
with respect to $X$), is non-monotonic in $X$ and there exists at least one 
inflection point $X_{i}$, where             
\begin{equation}                               
		       f\ ^{\prime\prime}(X_{i}, a, b)  = 0.                                                    
\end{equation}                                
Then we consider the following two cases.\\
Case (i)
\begin{equation}                           
		       f\ ^{\prime\prime\prime}(X_{i}, a, b)  > 0                                              
\end{equation}                           
We define a value of $a$ as $a_{1}$, through the relation,
\begin{equation}                             
		       f\ ^{\prime}(X_{i}, a_{1}, b) = -1, 
\end{equation}                      
the first period doubling point of the system. For a value of $a$ close to 
$a_{1}$ but greater than $a_{1}$, by adjusting the additive parameter $b$, the 
system can be taken through a bubble structure in the bifurcation scenario.\\
Case (ii)
\begin{equation}                         
			f\ ^{\prime\prime\prime}(X_{i}, a, b)  < 0                                              
\end{equation}                           
Here $a_{1}$ is defined as the tangent bifurcation point of the system, through 
the relation
\begin{equation}                             
		       f\ ^{\prime}(X_{i}, a_{1}, b) =+1 
\end{equation}                      
Then by fixing $a$ greater than $a_{1}$, but close to $a_{1}$, and tuning $b$, a 
bistability region can be produced in the system.\\
For case (i), the value of $a$ chosen to be greater than $a_{1}$ makes 
$f\ ^{\prime}(X_{i}, a, b)< -1$ or $\left | f\ ^{\prime}(X_{i}, a, b) \right | >1$. Moreover, 
conditions (2) and (3) imply that $X_{i}$ is a minimum for $f\ ^{\prime}$, which 
is concave upwards on both sides of $X_{i}$. Hence for a fixed point $X_{-}^{*}$ 
which is to the left of $X_{i}$, but in the immediate neighborhood, of $X_{i}$, 
$ \left | f\ ^{\prime}(X_{-}^{*}, a, b) \right | < 1$, and hence will be stable.Similarly, fixed 
point $X_{+}^{*}$ to the right of $X_{i}$, but near to $X_{i}$, 
$ \left | f\ ^{\prime}(X_{+}^{*}, a, b) \right | <1$ and is stable. Now, the second parameter $b$ 
is simple additive for the class of maps under consideration and hence 
$f\ ^{\prime}$ is independent of $b$. By adjusting $b$, the fixed point can be 
shifted such that $f\ ^{\prime}(X_{-}^{*}, a, b)$ becomes equal to -1, the period 
doubling point of the map.Then $X_{-}^{*}$ will give rise to a 2-cycle with 
elements $X_{1}^{*}$ and $X_{2}^{*}$. Since these are in the neighborhood of $X_{i}$, 
$f\ ^{\prime}(X_{1}^{*})$ and $f\ ^{\prime}(X_{2}^{*})$ will be negative so that 
the product $f\ ^{\prime}(X_{1}^{*})f\ ^{\prime}(X_{2}^{*})$ is positive. With further 
increase of $b$, $a$ period merging takes place for the 2-cycle, with $X_{1}^{*}$ 
and $X_{2}^{*}$ collapsing into $X_{+}^{*}$, which is just stable at the point 
where $f\ ^{\prime}(X_{+}^{*})=-1$. Thus in the parameter window $(b_{1}, b_{2})$, a 
bubble structure is formed.\\
The situation is exactly reversed for case (ii). Here conditions (2) and (5) 
makes $X_{i}$ a maximum of $f\,^{\prime}$ and the falls off on both sides of 
$X_{i}$. At a value of $a>a_{1}$, where $a_{1}$ is defined by (6), 
$f\ ^{\prime}(X_{i}, a, b)>+1$. Then in the neighborhood of $X_{i}$, a fixed point 
$X_{-}^{*}$, to the left of $X_{i}$, can be stable since 
$\left | f\ ^{\prime}(X_{i}, a, b) \right |<1$. 
Similarly $X_{+}^{*}$ on the right of $X_{i}$ also will be stable.By adjusting 
the second parameter $b$, these will be shifted to their respective tangent 
bifurcation points, i.e., $b_{1}$ where $X_{+}^{*}$ is born and $b_{2}$ where 
$X_{-}^{*}$ disappears. Then a bistability window is seen in the interval 
$(b_{1}, b_{2})$.\\
{\large \bf {3. Graphical analysis.}}\\
The mechanism of occurrence of bubbling and bistability explained above for 
maps satisfying the conditions in case (i) and case (ii) respectively can be 
made more transparent through a detailed graphical analysis. For this we 
plot the curve C1=$f\ ^{\prime}(X)$, the 1-cycle fixed point curve C2=${f(X)-X}$ 
and the 2-cycle curve C3=${f(f(X))-X}$ simultaneously as functions of $X$, for 
chosen values of $a$ and $b$. The zeroes of C2 give the 1-cycle fixed point 
$X^{*}$ while those of C3 give the elements of the 2-cycle. Their stability 
can be checked from the same graph, since the value of the derivative at the 
fixed points can be read off. We fix the value of $a$ to be greater than $a_{1}$, 
which helps to position the curve C1 in the proper way. By plotting the three 
curves for different values of $b$, bistability regions or bubbling sequences 
can be traced for any given map function of type (1).\\
For further discussion, we consider two specific forms of maps of the cubic 
type, which are simple but typical examples for case (i) and (ii). They are, 
\begin{equation}
M1 :       X_{n+1} = b - a X_{n} + X_{n}^ {3}                                            
\end{equation}
\begin{equation}
M2 :       X_{n+1} = b + a X_{n}  - X_{n}^ {3}                                            
\end{equation}
For M1, there are two critical points, $X_{c1}=-\sqrt{a/3}$, which is a maximum and 
$X_{c2}=\sqrt{a/3}$, which is a minimum. The inflection point in between 
occurs at $X_{i}=0$ and $f\ ^{\prime\prime\prime}=6$. Hence it belongs to case (i) 
and $a_{1}$ as defined by (4) is 1. In Fig(2), the three curves mentioned above 
are plotted for this map at $a=1.3$. We start from a value of $b = -1.34$, 
Fig(2.a) where the fixed point $X_{-}^{*}$ is just born via tangent 
bifurcation since $f\ ^{\prime}(X_{-}^{*})$ here is +1, and the curves C2 and C3 
just touches the zero line on the left of $X_{i}$ at $X_{-}^{*}$. Though C2 has 
a zero on the right, the slope there is larger than 1 and hence it is 
unstable, for this value of $b$. Since $b$ is only additive, increase in the  
value of $b$, shifts C2 upwards, resulting in a slow drift of $X_{-}^{*}$ from 
left to right. Thus as $b$ is increased to -0.7(Fig(2.b)), 
$f\ ^{\prime}(X_{-}^{*})=-1$ and $X_{-}^{*}$ bifurcates into $X_{1}^{*}$ and 
$X_{2}^{*}$. At $b= -0.3$(Fig(2.c)), the 2-cycle is stable with 
$f\ ^{\prime}(X_{1}^{*})$ and $f\ ^{\prime}(X_{2}^{*})$, both negative and their product 
is positive but less than 1. Note that the curve C3 has developed a maximum 
and a minimum on both sides of $X_{-}^{*}$, which is now unstable, cutting the 
zero line again at $X_{1}^{*}< X_{-}^{*}$ and $X_{2}^{*}>X_{-}^{*}$. As $b$ is further 
increased, they move apart. Since the value chosen is within the stability 
window of 2-cycle no further period doubling takes place. As $X_{-}^{*}$ 
crosses $X_{i}$, $X_{1}^{*}$ $\&$ $X_{2}^{*}$ move towards each other and merge together 
at $b=0.7$(Fig(2.d)) and coincide with the fixed point $X_{+}^{*}$. Further, 
$X_{+}^{*}$ disappears by a reverse tangent bifurcation at $b=1.34$, when 
$f\ ^{\prime}(X_{+}^{*})$ becomes equal to +1. Thus the above events lead to the 
formation of a primary bubble in the window (-0.7,0.7).\\
By keeping $a$ at a value beyond the second period doubling point $a_{2}$, of 
the map, the merging tendency starts only after the second period doubling 
and hence secondary bubbles are seen on the arms of the primary bubble. These 
can be repeated until at $a>a_{\infty}$, the system is taken to chaos.\\ 
Now the above analysis is repeated for map M2, which satisfies the conditions 
in case (ii) (Fig(3)). Here of the two critical points of the map, 
$X_{c1}= -\sqrt{a/3}$ is the minimum and $X_{c2}=\sqrt{a/3}$ is the maximum 
with a positive slope at the point of inflection $X_{i}$. $a_{1}$ in this case 
is also 1. Hence in the Fig (3) $a$ is chosen to be 1.4. Fig(3.a) shows 
the situation for $b= -0.35$, where $f\ ^{\prime}(X^{*})=-1$ and hence the 
1-cycle fixed point $X_{-}^{*}$ period doubles into a 2-cycle. For lower values 
of $b$, we expect the full period doubling scenario since $f\ ^{\prime}$ is 
monotonic beyond this point(Fig(3.b)). However, as $b$ is increased to -0.1, 
$f\ ^{\prime}(X_{+}^{*})=+1$, where the other 1-cycle, $X_{+}^{*}$ to the right of 
$X_{i}$ is born by tangent bifurcation(Fig(3.c)). Note that at this point $X_{-}^{*}$ is still 
stable with $\left | {f\ ^{\prime}(X_{-}^{*})}\right |<1$. This continues until $b= +0.1$, 
where $f\ ^{\prime}(X_{-}^{*})=+1$ and hence $X_{-}^{*}$ disappears(Fig(3.d)). The birth of 
$X_{+}^{*}$ is concurrent with the maximum of C2 touching the zero line $(b=b_{1})$  
while the disappearance of $X_{-}^{*}$ occurs as the minimum of C2 touches the 
zero line $(b=b_{2})$. As $b$ is increased and C2 is moving up it is clear that 
the former will take place for a lower $b$ value than the latter as the 
maximum of C2 occurs for $X>X_{i}$ and minimum at $X<X_{i}$ (slope 
being positive at $X_{i}$). Hence $b_{1}<b_{2}$, or there is a window $(b_{1}, b_{2})$, 
where bistability exists, which in our graph is (-0.1,0.1) for $a=1.4$. 
$X_{+}^{*}$ is stable beyond this point also and it period doubles as $b$ is 
increased to $b= +0.35$ where $f\ ^{\prime}(X_{+}^{*})=-1$. The full Feigenbaum 
scenario then develops for higher values of $b$.By keeping $a$ at higher 
values and tuning $b$, the bistability can be taken to 2-cycle, 4-cycle and 
even chaotic regions.\\ 
The stability regions of the different types of dynamical behaviour possible 
for M1 can be marked out in parameter space plot in the $(a, b)$ 
plane(Fig(4.a)). The cone like region on the left is the stability zone of the 
1-cycle fixed point (periodicity, p=1) and it is separated from the escape 
region by the tangent bifurcation line on both sides. The parabola like 
curve inside it marks out the 2-cycle (p=2) region, while the smaller 
parabolas indicate curves along which 4-cycles (p=4) and other higher 
periodic cycles becomes stable until chaos is reached. The line parallel to 
the $b$-axis at a value of $a>a_1$, along which primary bubble is formed, is 
shown by the dotted line. It is clear that along this  line, the system is 
taken from escape --> 1-cycle -->2-cycle -->1-cycle --> escape. Similarly 
secondary bubbles are formed along a line drawn at $a>a_2$ etc. The 
parameter space plot for M2 is shown in Fig(4.b). The quadrilateral like 
region marked as (I) beyond $a> a_{1}$ is the bistable region for 1-cycle, 
while quadrilateral (II) is that for 2-cycle etc. The area marked with p=1, 
is the stability region of 1-cycle while p=2, that for 2-cycle etc. When 
the system is taken along the dotted line beyond $a_{1}$, bistability is seen 
in the central region, followed by period doubling bifurcations to both sides, 
until chaos is reached.\\
{ \large \bf {4. Conclusion.}}\\
Although the above discussion is confined to two simple cubic maps, the 
analysis is repeated for a large number of maps of type (1) chosen from a 
wide variety of situations covering different functional forms like 
exponential, trigonometric and polynomial maps. We find that the qualitative behaviour in all cases remain the same 
and depends only on the criteria (2)-(6). Hence the pattern of scenario 
detailed in this paper can be taken to be atypical as far as maps of the 
form (1) are concerned.\\  
The criteria for bistability reported here are certainly novel while those 
for bubbling are more rigorous and general in nature compared to earlier 
studies. They can be used as a test to identify maps in which bistability or 
bubbling is possible and also to isolate the regions in the parameter space 
$(a, b)$ where they occur. Our main result is that whether it is bistability 
or bubbling is decided by the sign of the third derivative of the map 
function at the inflection point. If  $f\ ^{\prime\prime\prime}(X_{i})$ is 
positive, because of the concave nature of the derivative, tangent 
bifurcation will precede period doubling as $b$ is increased. Hence bubbling 
structure is possible. Similarly when $f\ ^{\prime\prime\prime}(X_{i})$ is 
negative, curve of $f\ ^{\prime}$ is convex and hence period doubling 
precedes tangent bifurcation, leading to bistability. In case  
$f\ ^{\prime\prime\prime}(X_{i})=0$, higher derivatives must be considered for 
deciding the behaviour.\\ 
Bubbling can be looked upon as an extreme case of incomplete period 
doublings and the latter has been often associated with positive Schwarzian 
derivative[22]. But for the system under study, it is easy to check that 
this is always negative (independent of the form of the map function), 
because of properties (2), (3) and (5). In fact, a few such 
maps have been reported earlier[23] though in a totally different context.
The bubbling scenario in maps of the type  M1, leads to bimodal chaos that 
is restricted to the arms of the primary bubble. Such confined chaos or even 
low periodic behavior prior to that, makes them better models in population 
dynamics of eco systems than the usual logistic type maps[24]. Attempts to 
extend the criteria to continuous and higher dimensional systems are 
underway and will be reported elsewhere.\\
{\large \bf {Acknowledgements}}\\
SNV thanks the UGC, New Delhi for financial assistance through a junior 
research fellowship and GA acknowledges the warm hospitality and computer 
facility at \\ IUCAA, Pune.\\

\pagebreak
{\large \bf {References}}\\
$\left [ 1 \right ]$ S Parthasarathy \& S Sinha, { \it Phys. Rev}{ \bf E51}, 6239(1995)\\
$\left [ 2 \right ]$ P R Krishnan Nair, V M Nandakumaran \& G Ambika,\\ { \it Pramana (J.Phys.)} 
{ \bf 43}, 421 (1994)\\
$\left [ 3 \right ] $ M Bier \& T Bountis, { \it Phys. Lett.} { \bf A104}, 239(1984)\\
$\left [ 4 \right ]$ G L Oppo \& A Politi, { \it Phys. Rev.} { \bf A30}, 435(1984)\\
$\left [ 5 \right ]$ C Lepers, J Legrand \& P Glorieux, { \it Phys.Rev.} { \bf A43}, 2573(1991)\\
$\left [ 6 \right ]$ T Hogg \& B A Huberman, { \it Phys. Rev.} { \bf A29}, 275(1984)\\
$\left [ 7 \right ]$ J Kozlowski, U Parlitz \& W Lauterborn, { \it Phys. Rev.} {\bf E51},1861(1995)\\
$\left [ 8 \right ]$ C S Wang, Y H Kao, J C Huang \& Y S Gou, { \it Phys.Rev.} {\bf A45}, 3471(1992)\\
$\left [ 9 \right ]$ K Coffman, W D McCormick \& H L Swinney, { \it Phys.Rev.Lett.} {\bf 56}, 999(1986)\\
$\left [ 10 \right ]$ T S Bellows, {\it J.Anim. Ecol.}, {\bf 50}, 139(1981)\\
$\left [ 11 \right ]$ M R Guevara, L Glass \& A Shrier, {\it Science}, {\bf 214}, 1350(1981)\\
$\left [ 12 \right ]$ P R Krishnan Nair, V M Nandakumaran \& G Ambika,{\it Computational Aspects 
in Nonlinear dynamics \& Chaos, eds.- G Ambika \& V M Nandakumaran (Wiley 
Eastern Pub.Ltd., New Delhi)},{\bf 144}, (1994)\\
$\left [ 13 \right ]$ P P Saratchandran,V M Nandakumaran \& G Ambika,{\it Pramana(J.Phys.)} 
{\bf 47}, 339(1996)\\
$\left [ 14 \right ]$ Z Qu, G Hu, G Yang \& G Qiu, {\it Phys. Rev. Lett.} {\bf 74}, 1736(1995)\\
$\left [ 15 \right ]$ C Uhl \& D Fournier-Prunaret, {\it Int.J.Bif.}\& {\it Chaos},{\bf 5}, 1033(1995)\\
$\left [ 16 \right ]$ X Zhang \& D F Jarett, {\it Chaos} {\bf 8}, 503(1998)\\
$\left [ 17 \right ]$ L Stone, {\it Nature}, {\bf 365}, 617(1993)\\
$\left [ 18 \right ]$ S Sinha \& P Das, {\it Pramana (J.Phys.)} {\bf 48}, 87(1997)\\
$\left [ 19 \right ]$ R Roy \& L Mahdel, {\it Opt.Commun.} {\bf 34}, 133(1980)\\
$\left [ 20 \right ]$ L O Chua \& K A Stromsmoe, {\it Int.J.of Engg.Science},{\bf 9}, 435(1971)\\
$\left [ 21 \right ]$ G Nicolis, C Nicolis \& D McKerman, {\it J.Stat.Phys.} {\bf 70}, 125(1993)\\
$\left [ 22 \right ]$ D Singer, {\it Int. J. Appl. Math.}{\bf 35}, 260(1978)\\
$\left [ 23 \right ]$ H E Nusse \& J A Yorke, {\it Phys. Rev. Lett.}{ \bf 27}, 328(1988)\\
$\left [ 24 \right ]$ S Sinha \& S Parthasarathy, {\it Proc.Natl.Acad. Sci.USA }{ \bf 93},1504(1996)\\

\pagebreak

{\large \bf {Figure Captions}}\\
Fig.1:- \hspace{.2 cm} Bifurcation diagram showing (a) bubble structure and 
(b) bistable behaviour, for a fixed value of $a$, with $b$ as the control 
parameter.

Fig.2:- \hspace{.2 cm} The derivative curve C1, the 1-cycle solution curve 
C2 and the 2-cycle solution curve C3 plotted  with the value of  $a$  at 1.3 
for the map M1. In (a),  $b = -1.34$ shows the point where the 1-cycle 
$X_{-}^{*}$ is just born, with $f\ ^{\prime}(X_{-}^{*})= +1$. (b) With 
$b= b_{1}= -0.7$, $f\ ^{\prime}(X_{-}^{*})= -1$ 
hence the $X_{-}^{*}$ becomes unstable and the 2-cycle is just born.(c) 
$b=-0.3$, shows the elements of the stable 2-cycle with $X_{1}^{*}$  to the 
left and $X_{2}^{*}$ to the right of the $X_{-}^{*}$, which is unstable now 
and (d) $b=b_{2}= +0.7$, the 1-cycle fixed point  $X_{+}^{*}$ becomes stable 
after the merging of $X_{1}^{*}$ and $X_{2}^{*}$.

Fig.3:- \hspace{.2 cm} Here the curves C1 ,C2 and C3 for the map M2 defined in (8) with 
$a= 1.4$ plotted. (a) At $b= -0.5$, it is clear from the figure that the 1-cycle 
solution is unstable and the 2-cycle is stable. (b) $b = -0.35$ gives the 
first period doubling point i.e., here $f\ ^{\prime}(X_{-}^{*})= -1$.   
(c) At $b = b_{1}= - 0.1$, $f\ ^{\prime}(X_{+}^{*})= +1$, i.e., the creation 
of a new fixed point $X_{+}^{*}$ by tangent bifurcation. Note that still 
$X_{-}^{*}$ is stable and (d) $b= b_{2}=0.1$, $f\ ^{\prime}(X_{-}^{*})$ is +1. 
Hence the existing fixed point $X_{-}^{*}$ disappears. Thus $(b_{1}, b_{2})$ 
gives the bistability window.

Fig.4:-\hspace{.2 cm} Parameter space plot in $(a, b)$ plane (a) for map M1 
and (b) for map M2. 

\pagebreak
\end{document}